# A New Method for Epileptic Seizure Classification in EEG Using Adapted Wavelet Packets

Amirmasoud Ahmadi
Biomedical Engineering Department
School of Electrical Engineering, Iran
University of Science and Technology
Tehran, Iran
amirmasoud_ahmadi@elec.iust.ac.ir

Vahid Shalchyan
Biomedical Engineering Department
School of Electrical Engineering, Iran
University of Science and Technology
Tehran, Iran
shalchyan@iust.ac.ir

Mohammad Reza Daliri
Biomedical Engineering Department
School of Electrical Engineering, Iran
University of Science and Technology
Tehran, Iran
daliri@iust.ac.ir

*Abstract*—Electroencephalography (EEG), as the most common tool for epileptic seizure classification, contains useful information about different physiological states of the brain. Seizure related features in EEG signals can be better identified when localized in time-frequency basis projections. In this work, a novel method for epileptic seizure classification based on wavelet packets (WPs) is presented in which both mother wavelet function and WP bases are adapted a posteriori to improve the seizure classification. A support vector machine (SVM) as classifier is used for seizure versus non-seizure EEG segment classification. In order to evaluate the proposed algorithm, a publicly available dataset containing different groups' patient with epilepsy and healthy individuals are used. The obtained results indicate that the proposed method outperforms some previously proposed algorithms in epileptic seizure classification.

## I. Introduction

Electroencephalogram (EEG) is the most common methods for investigating a seizure disorder by detecting abnormal electrical activity in the brain through small electrodes applied over the scalp surface[1, 2]. Since the treatment of seizures depends on an accurate diagnosis, it is important to make sure that a patient has epilepsy and knowing what kind of disorder it has [3, 4]. Many feature extraction algorithms have been developed for seizure detection, based on time domain [6], frequency domain[7, 8], and time-frequency domain [9-12] analysis. Since the EEG is considered as a non-stationary signal, among feature extraction methods, time-frequency based methods such as wavelet transforms are usually better choices to process these kinds of signals because those can better reflect and localize the time-varying frequency characteristics of the data. Wavelet packets transform (WPT) is a powerful method that can even better localize the important signal feature by selecting the best basis for transformation [9, 13].Feature selections which can effectively reveal the characteristic of the original EEG signal are very important. The selection can be used to assess the effectiveness of the wavelet coefficients [5, 9]. Various linear and non-linear time-domain features have been used in the published articles included Lyapunov exponents[14], entropy [11, 15], energy [16, 17], minimum peak, maximum peak, mean value, standard deviation[10, 17-19], and root mean square (RMS) [20].

There are also a wide variety of classification methods for seizure type prediction such as artificial neural networks (ANN) [15, 21], support vector machine (SVM) [14,15, 16], and other machine learning methods [18, 22]. SVM is a perfect tool for classification [23] and widely used for seizure detection and classification [12, 18, 22, 24].

In this study, we presented a novel algorithm based on WPT combined with various features for the feature extraction and SVM as classifier for seizure classification problems.

## II. MATERIAL AND METHODS

The proposed algorithm is shown as a block diagram in Fig.2. Each EEG data was divided into 17 sub-segments. The sub-segments were decomposed to WP trees with different mother wavelets. The features were extracted from the best mother wavelets and WP bases and were used by a SVM classifier.

### A. EEG data

We used the EEG data collected from the Epilepsy Centre at the Bonn University [25]. Five sets of the EEG data, A, B, C, D, and E, each containing 100 single channel EEG segments of 23.6 seconds duration were used. Recordings were from 10 subjects, five healthy individuals and five patients with epilepsy. From this set, A and B were considered as non-seizure conditions and C, D, and E, were considered as seizure cases. The EEG datasets A (Set A) and B (Set B) were gotten from healthy individuals with eyes open and closed, respectively; sets C and D were gotten during seizure free interval in different zone of the brain and set E was obtained from a patient during ictal period [25](Table I). In order to obtain an appropriate segment of the database, all segments of EEG signal of each set were divided into 17 sub-segments (1.38 seconds) generating 1700 samples from each set; for example, set A was divided to $a_{1-1} \cdots a_{1-17}$.

TABLE I . Summary of datasets obtained from EEG database of University of Bonn (UoB), Germany [30]

| Subject information | Five healthy subjects | | Five epilepsy patients | | |
|---|---|---|---|---|---|
| | SET A | SET B | SET C | SET D | SET E |
| Patient state | Awake with eyes open (normal) | Awake with eyes close (normal) | Seizure free (interictal) | Seizure free (interictal) | Seizure activity (ictal) |
| Electrode type | Surface | Surface | Intracranial | Intracranial | Intracranial |
| Electrode placement | International 10–20 system | International 10–20 system | Opposite to epileptogenic zone | Within epileptogenic zone | Within epileptogenic zone |
| No. of epochs | 100 | 100 | 100 | 100 | 100 |
| Epochs duration | 23.600 s | 23.600 s | 23.600 s | 23.600 s | 23.600 s |

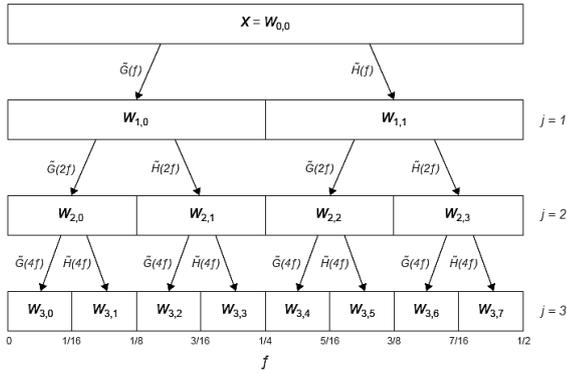

**Fig 1:** Wavelet packet decomposition tree

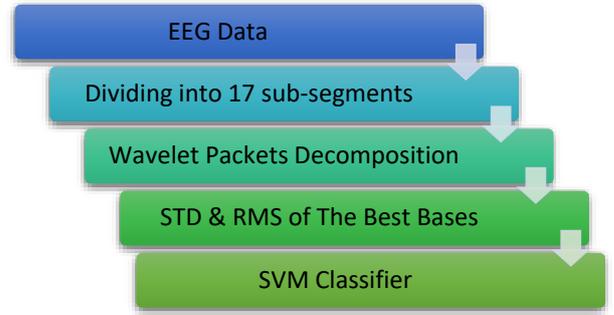

**Fig 2:** Block diagram of the proposed algorithm.

$$A_1 = \begin{bmatrix} a_{1-1} \\ \vdots \\ a_{1-17} \end{bmatrix} \cdots A_{100} = \begin{bmatrix} a_{100-1} \\ \vdots \\ a_{100-17} \end{bmatrix}. \quad (1)$$

### B. Feature Extraction

Wavelet transform has been developed as an important tool in feature extraction of nonstationary signals. Wavelet transform has better resolution and higher performance in determining epileptic seizure activity compared to the short-time Fourier transform (STFT) [26]. In this work we used wavelet packets transform for as a tool for selecting best bases for the feature extraction. Wavelet packets transform generates a multi-level symmetric decomposition tree (Fig. 1). In each decomposition level all sub-bands symmetrically decomposed into high and low-pass filtered components both of which are projections onto orthogonal basis functions [27]. WPT unlike WT decomposes both approximation and detail at each level, and has the ability to localize important information in the high frequency components.

Statistics over the set of the wavelet packets coefficients are used to further reduce the feature vectors [19]. The statistical features which were used in this paper are the standard deviation (STD) and the root-mean-square (RMS) of WP coefficients. We decomposed each of the signals (1.36 seconds) into WPT components over five levels with different mother wavelet functions including db2, sym4, rbio2.2, db6, bior2.4, bior1.1 as were mostly used in previous studies [11, 12, 19].

Knowing that the seizure in recorded EEGs usually occurs between 3-29 Hz [40] the decomposition level was set to five and the candidate WP bases were primarily selected to match well with the aforementioned frequency range of the EEGs signals. Best mother wavelet and best bases over all candidate WPT bases were determined according to the posterior classification performances obtained as shown in the results section (Table III, Fig.3).

### C. Classification

Support vector machines with different kernel functions are widely used for pattern classification [5]. In the current study, we used SVMs for different binary classification problems with a radial basis function (RBF) kernel. (Fig.2). Seven case of classification were considered: *A versus E, B versus E, C versus E, D versus E, AB versus E, CD versus E, ABCD versus E*.

### D. Evaluation

In order to evaluate the proposed algorithm, a k-fold cross-validation paradigm was used. The data (i.e., 1700 samples) in each group were divided into k-folds (k= 2, 5, 10).

## III. RESULT

### A. Selecting the best WP bases

Table II compares the effect of WP bases selection on the classification accuracy (CA) for ABCD vs E classification based on a 2-fold cross validation. The results showed that the WP coefficients (5,1), (4,1), (4,2) performed better than the other WP basis candidates. Therefore; the WP coefficients (5,1), (4,1), (4,2) were fixed as the best WP bases for all other analyses in this work.

### B. Selecting the best mother wavelet

Fig. 3 compares the effect of selecting different mother wavelets on CA performance for ABCD vs. E classification based on a 2-fold cross validation. The obtained result showed that the Bior1.1 outperformed other candidate mother wavelets. Based on this results, the Bior1.1 was selected as the best mother wavelet for all other analyses in this work.

TABLE II. CA performance of ABCD vs E, 2-fold, mother wavelet Db4

| WP Coefficient | CA% | WP Coefficient | CA% |
|---|---|---|---|
| (5,1) (4,1) (4,2) | **93.5%** | (5,1) (4,1) (4,2) (4,3) | 90% |
| (5,1) (4,1) (3,1) | 89% | (5,0] (5,1) (4,1) (4,2) | 89% |
| (4,1) (4,2) (4,3) | 91% | (5,0) (5,1) (4,1) | 91% |

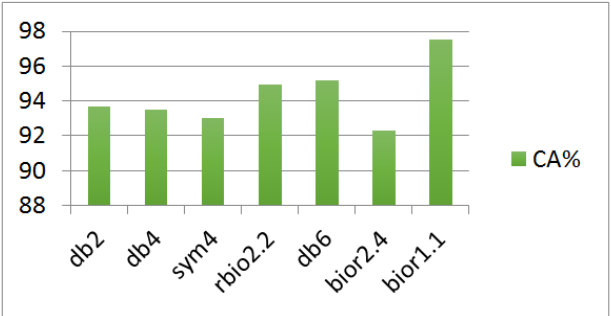

Fig 3: CA performance for different mother wavelets

### C. Seizure versus non-seizure

Table III describes all classification performance measures for seven different classification problems, described in the method part. As shown in the table, the CA performance varies between 96.04% to 99.64% each corresponds to the cases A versus E and D versus E respectively. Tables IV compares the CA performance of the proposed methods with some of the previously published methods in cases A versus E, B versus E, C versus E, D versus E and ABCD Versus E.

TABLE III. Results obtained after classification of seizure versus non-seizure

| Case | Fold | Performance | | |
|---|---|---|---|---|
| | | CA | Sens | Spec |
| A Versus E | 2-fold | 99.64% | 99.70% | 100% |
| | 5-fold | 99.64% | 99.64% | 100% |
| | 10-fold | 99.64% | 99.64% | 100% |
| B Versus E | 2-fold | 98.41% | 98.47% | 99.94% |
| | 5-fold | 98.38% | 98% | 99.94% |
| | 10-fold | 98.44% | 98.05% | 99.94% |
| C Versus E | 2-fold | 98% | 98.05% | 99.29% |
| | 5-fold | 98.14% | 98.29% | 98.64% |
| | 10-fold | 98.14% | 98.17% | 98.58% |
| D Versus E | 2-fold | 94.86% | 97.74% | 95.18% |
| | 5-fold | 95.06% | 97.69% | 94.72% |
| | 10-fold | 95.15% | 97.42% | 94.89% |
| AB Versus E | 2-fold | 98.83% | 99.37% | 99.82% |
| | 5-fold | 98.96% | 98.91% | 99.80% |
| | 10-fold | 98.93% | 98.86% | 99.74% |
| CD Versus E | 2-fold | 96.04% | 98.29% | 96.23% |
| | 5-fold | 96.41% | 98.02% | 96.38% |
| | 10-fold | 96.48% | 97.94% | 96.35% |
| ABCD Versus E | 2-fold | 97.48% | 98.14% | 97.48% |
| | 5-fold | 97.63% | 98.24% | 97.47% |
| | 10-fold | 97.85% | 98.19% | 97.58% |

TABLE IV. Comparison of the proposed algorithm with other previous works from literature for classifications of A vs. E, B vs E, C vs E, D vs E and ABCD vs E in terms of CA performance.

| Case | Method | CA |
|---|---|---|
| A Vs E | Non-linear preprocessing filter + Neural networks [28] | 97.2 |
| | Time–frequency features + Recurrent neural networks [6] | 99.6 |
| | Entropy + Adaptive neuro-fuzzy inference system [29] | 92.22 |
| | FFT-decision tree classifier [30] | 98.72 |
| | Discrete wavelet transform, mixture of expert model [5] | 95 |
| | **Our Algorithm** | **99.64** |
| B Vs E | Technique-based least square support vector machine [31] | 93.6 |
| | Weighted horizontal visibility graph [32] | 97 |
| | Weighted visibility graph + Complex network feature [26] | 97.25 |
| | **Our Algorithm** | **98.44** |
| C Vs E | Weighted horizontal visibility graph [32] | 97 |
| | Technique-based least square support vector machine [31] | 98 |
| | Weighted visibility graph + Complex network feature [24] | 98.25 |
| | **Our Algorithm** | **98.14** |
| D Vs E | Technique-based least square support vector machine [31] | 93.6 |
| | Weighted horizontal visibility graph [32] | 93 |
| | Weighted visibility graph + Complex network feature [24] | 93.25 |
| | Permutation Entropy and Support Vector Machine [33] | 83.13 |
| | DWT based fuzzy approximate entropy + support vector machine [34] | 93 |
| | **Our Algorithm** | **95.15** |
| ABCD Vs E | Time-frequency analysis, artificial neural network [35] | 97.73 |
| | Time frequency analysis and + Linear least squares, linear [36] | 97.21 |
| | Wavelet transform + K-nearest neighbor classifier [37] | 97.41 |
| | Wavelet transform &Approximate + Artificial neural network [20] | 96.87 |
| | Wavelet transform &line length feature + Artificial neural network [39] | 97.17 |
| | **Our Algorithm** | **97.85** |

## IV. CONCLUSION

In this work, we used a wavelet packet scheme with selection of best mother wavelet function and best WP bases for extracting time-frequency feature of EEG signals for seizure prediction. The classification was done by SVM. The obtained results (Table IV) revealed that the proposed algorithm could effectively detect seizure. Compared to the previous studies, the proposed method outperformed previous methods in different binary classification scenarios including A vs. E ([5, 6, 28-30]), B vs. E ([24, 31, 33]), C vs. E ([31, 33]), D vs. E ([24, 31-34]), ABCD vs. D ([36-39]).